\documentclass[lettersize,journal]{IEEEtran}
\usepackage{amsmath,amsfonts,bbm}
\usepackage{algorithm}
\usepackage{algorithmic}
\usepackage{array}
\usepackage[caption=false,font=normalsize,labelfont=sf,textfont=sf]{subfig}
\usepackage{cite}
\usepackage{textcomp}
\usepackage{stfloats}
\usepackage{url}
\usepackage{verbatim}
\usepackage{graphicx}
\usepackage{stfloats}
\usepackage{color}

\def\BibTeX{{\rm B\kern-.05em{\sc i\kern-.025em b}\kern-.08em
    T\kern-.1667em\lower.7ex\hbox{E}\kern-.125emX}}
\usepackage{balance}

\usepackage[normalem]{ulem}

\begin{document}
\title{Integrated Sensing and Communication Driven Digital Twin for Intelligent Machine Network}
\author{Zhiqing Wei,~\IEEEmembership{Member,~IEEE}, 
Yucong Du,~\IEEEmembership{Student Member,~IEEE}, 
Qixun Zhang,~\IEEEmembership{Member,~IEEE},
Wangjun Jiang,~\IEEEmembership{Student Member,~IEEE},
Yanpeng Cui,~\IEEEmembership{Student Member,~IEEE},
Zeyang Meng,~\IEEEmembership{Student Member,~IEEE},\\
Huici Wu,~\IEEEmembership{Member,~IEEE}, 
and Zhiyong Feng,~\IEEEmembership{Senior Member,~IEEE}
\thanks{Z. Wei, Y. Du, Q. Zhang, W. Jiang, Y. Cui, Z. Meng, H. Wu and Z. Feng are with Key Laboratory of Universal Wireless Communications, Ministry of Education, Beijing University of Posts and Telecommunications. 
\emph{Corresponding author: Zhiyong Feng (e-mail: fengzy@bupt.edu.cn).}}}

\markboth{}
{}

\maketitle

\begin{abstract}
Intelligent machines (IMs), 
including industrial machines, 
unmanned aerial vehicles (UAVs), and unmanned vehicles, etc., 
could perform effective cooperation  
in complex environment when they form IM network. 
The efficient environment sensing and communication are 
crucial for IM network,
enabling the real-time and stable control of IMs. 
With the emergence of integrated sensing and communication (ISAC) technology,
IM network is empowered with ubiquitous sensing capabilities,
which is helpful in improving the 
efficiency of communication and sensing 
with the mutual benefit of them.
However, the massive amount of sensing information 
brings challenges for the processing, 
storage and application of sensing information.
In this article, 
ISAC driven digital twin (DT) is proposed
for IM network,
and the architecture and enabling technologies are revealed.
ISAC driven DT structurally stores the sensing information, 
which is further applied to optimize communication, networking
and control schemes of IMs,
promoting the widespread applications of IMs.
\end{abstract}


\section{Introduction}
Intelligent machines (IMs) \cite{1} are 
the machines
capable of performing tasks autonomously or interactively 
in complex environment, 
such as industrial machines, unmanned aerial vehicles (UAVs), 
unmanned vehicles, etc. 
Multiple IMs perform effective cooperation 
when they form IM network. 
In IM network, efficient environment sensing
and communication are essential for the 
real-time and stable control of IMs.
With the emergence of 
integrated sensing and communication (ISAC) technology
\cite{3},
IM network has potential to be 
empowered with ubiquitous sensing capabilities.
Exploiting the mutual benefit of sensing and communication,
the sensing accuracy and coverage are enhanced via
cooperative sensing.
Meanwhile, with the assistance of sensing information,
the communication delay is reduced and 
the efficiency of 
networking is improved.

Although ISAC brings advantages for IM network,
there are the following challenges 
in the ISAC-enabled IM network.
Firstly, ISAC-enabled IM network 
faces challenge in processing, storing and application of 
the large amount of sensing information.
Secondly, ISAC brings difficulty in 
the management of multi-domain
resources including communication, sensing, and computing,
which is difficult in supporting
low-cost and real-time control of IMs.
Finally, ISAC empowers IM network
with dual-functions of sensing and communication,
as well as network vulnerability.
Hence, the reliability enhancement of IM network
is challenging.

Facing the above challenges, 
the ISAC driven Digital Twin (DT) 
is applied in IM network connecting physical and digital spaces,
which has the following advantages. 
Firstly, ISAC driven DT aggregates distributed and 
centralized computing power 
in IM network, supporting the extraction, transmission, 
and decoding of 
semantic features in environment sensing information,
which constructs physical and wireless environment 
in the digital space.
Secondly, DT connects multi-domain resources 
of communication, sensing, and computing, 
enabling efficient, low-delay, and low-cost resource scheduling,
which supports the matching between 
differentiated IMs and diversified services.
Finally, DT empowers IM network with the capability of 
anomaly detection, 
enhancing the reliability of IM network.

Although the ISAC driven DT (ISAC-DT) 
enabled IM network has the above benefits,
it faces the following difficulties.
\begin{itemize}
	\def\labelenumi{\arabic{enumi})}

        \item \textbf{Environment sensing:}        
        The optimization of environment sensing schemes
        in highly dynamic environment is challenging.
        The large amount of diverse sensing information 
        brings challenges for the sensing information transmission and 
        processing.
        Besides, the fusion of multiple local DTs to 
        generate the global DT is challenging.
	
        \item \textbf{IM communication:}        
         The channel, node distribution and 
         interference in IM network are rapidly changing,
         the collection of physical layer data
         in DT construction is challenging and 
         the data may not be fresh,
         which degrades the optimization effect of 
         channel estimation and multiple access with the 
         assistance of ISAC-DT.
         
	\item \textbf{IM networking:} 
        The high mobility and multi-domain heterogeneous resources of IMs
        bring challenges to the real-time IM networking.
        The disguising nodes and malicious attacks bring 
        challenges to the reliability assurance of IM network.        
\end{itemize}

Facing the above challenges of ISAC-DT enabled 
IM network, there are some related studies. 
Cui \emph{et al.} \cite{16} established 
a physical layer architecture 
for ISAC-DT and studied the interference management and 
ISAC signal optimization schemes. 
Li \emph{et al.} 
\cite{17} optimized the DT-enabled integrated sensing, 
communication, and computing (ISCC) network. 
Gong \emph{et al.} \cite{18} studied 
task scheduling and resource allocation
in DT-enabled Internet of Vehicles (IoV). 
Although there are initial related works, 
they are still insufficient in addressing the 
above challenges of ISAC-DT enabled 
IM network. 
ISAC-DT assisted environment sensing and 
IM communication are rarely studied.
Besides, the ISAC-DT enabled real-time and reliable IM networking
are not sufficiently studied.
In this article, the architecture 
of ISAC-DT enabled IM network
is proposed, which could be applied in the optimization
of environment sensing, IM communication, 
IM networking, and IM control.
The enable technologies of ISAC-DT enabled IM network
including ISAC-DT based environment sensing,
ISAC-DT based IM communication and 
ISAC-DT based IM networking are studied.
Table \ref{tab1} provides a list of abbreviations in this article.

\begin{table}[]
\centering
\caption{Abbreviation Table.}
\label{tab1}
\begin{tabular}{|c|c|}
\hline
Abbreviation  & Full Name                                                                                           \\ \hline
6G            & 6th Generation                                                                                      \\ \hline
AGV           & Automated Guided Vehicle                                                                            \\ \hline
BS            & Base Station                                                                                        \\ \hline
CRA  & Complete Random Algorithm                                                                  \\ \hline
CSI           & Channel State Information                                                                           \\ \hline
DT            & Digital Twin                                                                                        \\ \hline
DTN           & Digital Twin Network                                                                                \\ \hline
ID            & Identification                                                                                      \\ \hline
IM            & Intelligent Machine                                                                                 \\ \hline
IP            & Internet Protocol                                                                                   \\ \hline
ISAC          & Integrated Sensing and Communication                                                                \\ \hline
ISAC-DT       & \begin{tabular}[c]{@{}c@{}}Integrated Sensing and Communication \\ Driven Digital Twin\end{tabular} \\ \hline
ISCC          & Integrated Sensing, Communication and Computing                                                     \\ \hline
IoV           & Internet of Vehicles                                                                                \\ \hline
LiDAR         & Laser Radar                                                                                         \\ \hline
MAC           & Medium Access Control                                                                               \\ \hline
MEC           & Mobile Edge Computing                                                                               \\ \hline
mmWave        & millimeter-Wave                                                                                     \\ \hline
OFDM & Orthogonal Frequency Division Multiplexing                                           \\ \hline
PLC           & Programmable Logic Controller                                                                       \\ \hline
RSU           & Road Side Unit                                                                                      \\ \hline
SLAM          & Simultaneous Localization and Mapping                                                               \\ \hline
SNR  & Signal-to-Noise Ratio                                                                     \\ \hline
THz           & Tera Hertz                                                                                          \\ \hline
UAV           & Unmanned Aerial Vehicle                                                                             \\ \hline
UI            & User Interface                     
                                                 \\ \hline
\end{tabular}
\end{table}

\section{Scenarios of ISAC-DT enabled IM Network}

\begin{figure*}
	\centering
	\includegraphics[width=0.99\linewidth]{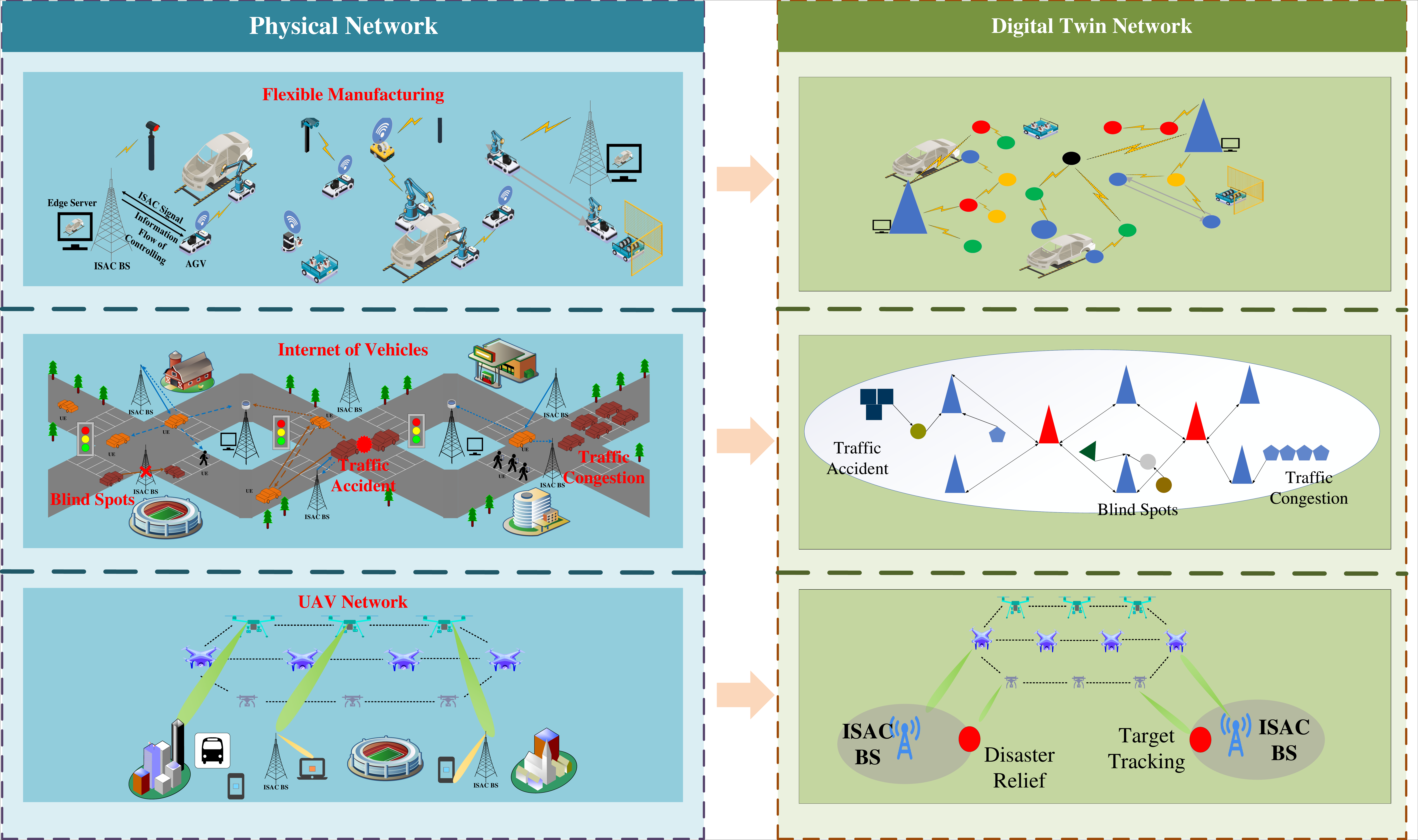}
	\caption{Scenarios of IM network, including flexible manufacturing, IoV, and UAV network.}
	\label{fig1}
\end{figure*}

The typical scenarios of IM network include 
flexible manufacturing, 
IoV
and UAV network, 
as shown in Fig. \ref{fig1}.

Flexible manufacturing enables 
the reconstruction of production lines according to 
differentiated service requirements,
which consists of Base Stations (BSs) equipped with Mobile Edge Computing (MEC) server and  
IMs such as Automated Guided Vehicles (AGVs), 
robotic arms, etc.
The IMs are controlled through by the 
Programmable Logic Controllers (PLCs) connecting with BSs.
The BSs and IMs with ISAC capability 
detect target and reconstruct factory environment.
Cooperative sensing among multiple BSs and between the 
BSs and IMs are performed to realize 
high-accuracy and large-coverage sensing.
Meanwhile, the sensing information is applied in 
constructing real-time DT of factory, 
which is further applied in the optimization of  
sensing, communication, networking and control schemes of IMs.

IoV consists of the Road Side Units (RSUs) and the IMs
such as vehicles. 
With RSUs and vehicles empowered with ISAC capability,
the environment sensing information is
aggregated to construct the DT revealing 
the real-time information of road, humans and vehicles,
which provides the information on traffic congestion, 
accidents, and blind spots for vehicles to enable the 
optimization of IoV and the control of traffic flow.

UAV network
consists of BSs and UAVs.
UAVs perform the tasks such as environment monitoring,
disaster management, etc.
The BSs with ISAC capability sense the environment and UAVs.
Then, the sensing information is sharing with UAVs to 
ensure the safety of UAVs.
Meanwhile, the UAVs with 
ISAC capability sense the environment and other UAVs in 
a more flexible way, 
compensating the blind areas of BS sensing.
The DT is constructed with the sensing information,
enabling the efficient environment sensing,
communication, networking and control of UAVs.

\section{Architecture of ISAC-DT enabled IM Network}
\label{Framework}

In this section, 
an architecture of ISAC-DT enabled IM network is proposed 
based on the architecture of 
digital twin network (DTN) proposed by 
China Mobile \cite{6, 19},
where the architecture of DTN \cite{6} 
consists of three layers
and two close-loops.
In ISAC-DT enabled IM network,
wireless and physical environment sensing information
is applied in building DT,
and the sensing information in DT 
assists environment sensing, IM communication and networking.
In the following subsections, 
the functions in each layer, 
module, and closed-loop are introduced. 

\begin{figure*}
	\centering
	\includegraphics[width=0.99\linewidth]{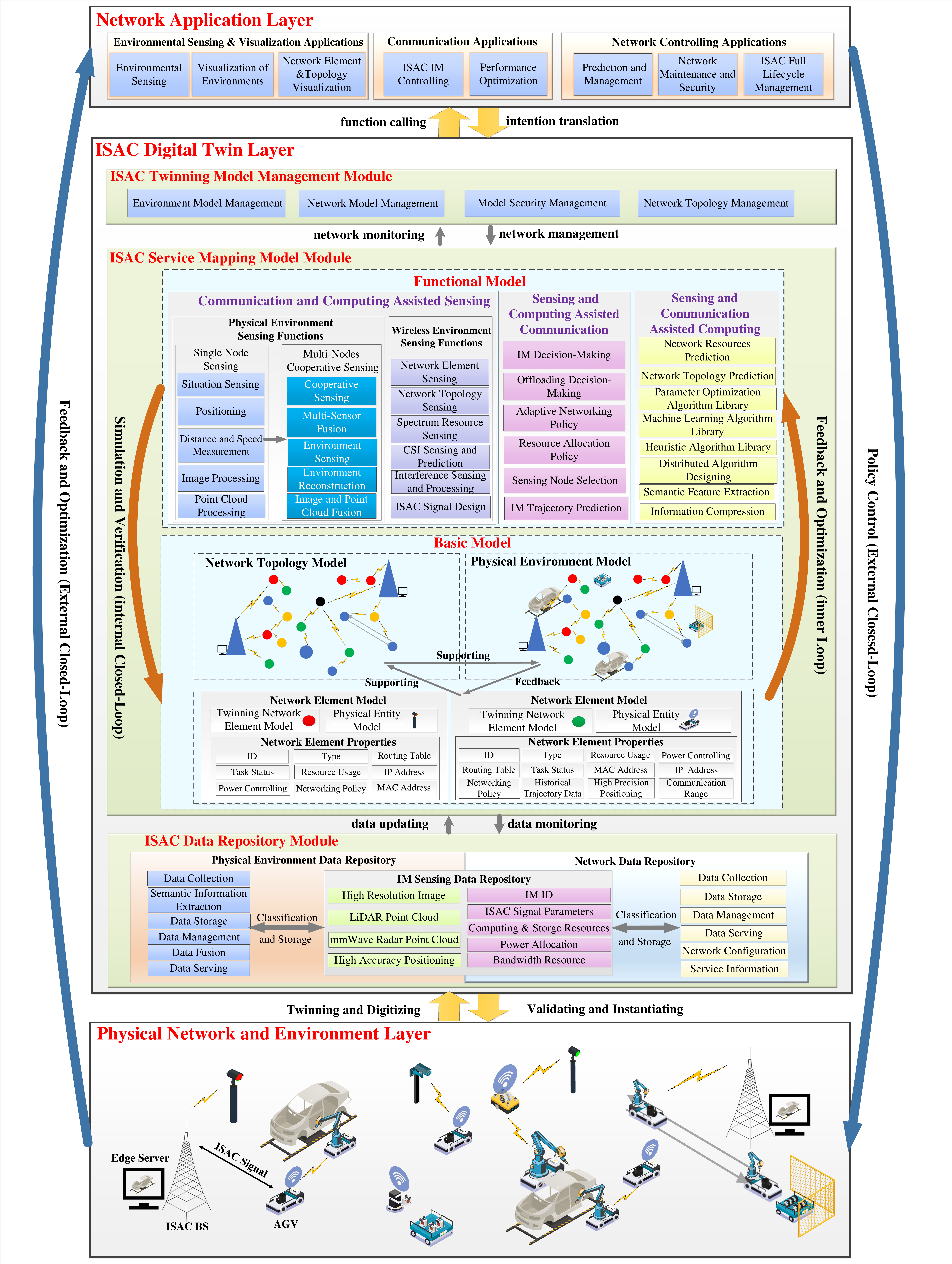}
	\caption{Architecture of ISAC-DT enabled IM network.}
	\label{fig2}
\end{figure*}

\subsection{Three Layers of ISAC-DT enabled IM Network}
The three layers of ISAC-DT enabled IM network 
are physical network and environment layer, 
ISAC digital twin layer, 
and network application layer.

\subsubsection{Physical Network and Environment Layer}
Physical network and environment layer 
aggregates the information of physical network and environment. 
After the simulating and validation in ISAC digital twin layer, 
policies are implemented in 
physical network and environment layer.

In contrast to the existing DTN,
which focus on the communication functions and services,
ISAC-DT enabled IM network focuses 
both on sensing and communication, 
as well as their mutual benefits. 
Hence, the physical network and physical environment
are aggregated into the physical network and environment layer .

\subsubsection{ISAC Digital Twin Layer}
ISAC digital twin layer 
maps the physical network and environment 
to twinning models, 
then simulating and verifying 
the policies according to the service requirements.
The detailed composition and 
functions of ISAC digital twin layer 
are explained in Section \ref{modules}.

\subsubsection{Network Application Layer}
The network application layer is connected 
with the services of IMs and users, 
and visualize the wireless and physical environments, network models and policy simulation results to IMs and users. 
Users can view the the real-time states of the network and environment through software or User Interface (UI).
Through intents (defining the goals or results of services or the network), users can participate or influence the policy making process.
compared to existing DTNs, 
the network application layer consists of sensing services.

\subsection{Three Modules in ISAC Digital Twin Layer}
\label{modules}
ISAC digital twin layer is the main entity of 
DT construction in the ISAC-DT enabled IM network,
which consists of three modules,
including ISAC data repository module, 
ISAC service mapping model module and 
ISAC twining model management module.

Compared with the existing DTNs, 
the wireless and physical environments are modeled
in the basic model.
Then, the wireless and physical environment 
sensing functions are provided 
in the functional model.
The mutual benefits of 
communication and sensing are considered 
in the functional model.

\subsubsection{ISAC Data Repository Module}
ISAC data repository module collects, 
separates and 
stores the heterogeneous sensing information and 
network information 
from internal and external sensors of IM network
into physical environment data repository 
and network data repository.
which assists the construction of the basic model, 
further supporting the simulation of 
various functions and algorithms 
in ISAC digital twin layer.

Network data repository manages the 
network configurations and services.
In physical environment data repository, 
the large amount of sensing information
is collected from various sensors
including camera, LiDAR and ISAC equipment, etc.
Then, semantic extraction and fusion 
of these sensing information are performed
in the functional model
and stored in physical environment data repository.

\subsubsection{ISAC Service Mapping Model Module}
ISAC service mapping model module enables 
policy making, validation, and management 
using the data in the ISAC data repository module, 
which consists of basic model and functional model. 

\paragraph{Basic Model}
Basic model is located between 
the ISAC data repository module and the functional model, 
which is constructed by 
unity software or existing DT platforms
launched by institutions, 
such as ONE TOTAL TWIN by Altair, Digital Twin Exchange by IBM, and Azure Digital Twins by Microsoft. 
In ISAC-DT enabled IM network, 
basic model including network element model, 
topology model, 
and environment model 
is constructed by the data in ISAC data repository.
Basic model is used for policy making 
in the functional model.
The details of network element model, 
topology model, and environment model are
revealed as follows.

\begin{itemize}
    \item \textit{Network Element Model} 
    contains the basic information of IMs, 
    including identification (ID), Internet Protocol (IP), type, Medium Access Control (MAC), routing table, 
    task status, resource utilization, etc..
    The Network Element Model can be deployed at the edge layer MEC servers,
    where the construction, storage and updating 
    of the network element model are performed 
    according to the physical environment sensing information 
    in the ISAC data repository module. 
    Network element model provides information for 
    network topology construction and updating,
    as well as provides support for the visualization function in the network application layer.

\item \textit{Network Topology Model} 
constructs and updates the route tables, 
spectrum utilization information and 
channel state information (CSI) of IM network. 
Network topology model could be constructed 
using network element information, 
such as routing table 
in the network element model and 
CSI of IM network. 
Network topology model
can also be constructed
using the prediction information 
in the functional model.

\item \textit{Physical Environment Model} 
fuses the network element model, 
network topology model and 
environment sensing information,
including all the information of 
IM network and flexible manufacturing factory.
Physical environment model can be deployed 
in the edge layer establishing local DT
and in the cloud layer establishing global DT.
Physical environment model provides digital network and environment 
for the simulation and validation in the functional model.
\end{itemize}

\paragraph{Functional Model}
Functional model is applied in 
policy making and verification
in cloud and edge layers,
which simulates the policies based on  
service requirements and time-varying basic model, 
while continuously updates algorithm parameters 
through internal close-loop to guarantee the real-time  
optimal performance of policies.
The functional model consists of three functions: 
communication and computing assisted sensing, 
sensing and computing assisted communication, 
and sensing and communication assisted computing,
which are revealed as follows.

\begin{itemize}
    \item \textit{Communication and computing assisted sensing:} 
    With the assistance of communication and 
    computing power in DT, 
the cooperative sensing is realized with the optimized
ISAC signals and node deployment. 

\item \textit{Sensing and computing assisted communication:}
With the assistance of sensing and computing power in DT, 
IMs obtain the optimal communication and 
networking schemes
in complex and dynamic environment.

\item \textit{Sensing and communication assisted computing:},
with the assistance of sensing and communication, 
the data is accumulated in DT, 
ensuring continuous update of algorithm libraries,
further increasing the effectiveness of prediction 
and policy making in DT. 
\end{itemize}

\subsubsection{ISAC Twining Model Management Module}
The ISAC twining model management module 
includes the following two functions. 
1) model management,
which includes the management of basic and functional models. 
2) model security and network reliability management,
which includes model encryption and security, 
as well as disguised nodes and malicious attack detection.

\subsection{Two Closed-loops of ISAC-DT enabled IM Network}
There are two closed-loops in the architecture of 
ISAC-DT enabled IM network. 
The internal closed-loop verifies 
the simulation and optimizes iteratively 
between basic model and functional model 
in the ISAC service mapping model module. 
The external closed-loop involves all layers 
in the ISAC-DT enabled IM network, 
the functions of which include policy control, 
feedback and optimization. 

\section{Enabling Technologies} \label{DT}

In this section, 
the enabling technologies of ISAC driven DT for 
IM network are revealed, 
including ISAC-DT based environment sensing,
ISAC-DT based IM communication, 
and ISAC-DT based IM networking.

\subsection{ISAC-DT based Environment Sensing}\label{3.2}
The ISAC-DT based sensing technologies enable efficient and intelligent sensing in IM network. 
However, they faces the following challenges.
Firstly, 
adaptive dynamic environment brings challenges of sensing optimization.
Secondly, 
the various information types and large amount of data brings challenges in sensing information transmission and processing for environment sensing.
Finally,
cooperative sensing of multi-BSs brings challenges in the fusion of multi-DTs in BSs.
Facing with the above challenges, we propose the following enabling technologies.

\subsubsection{Wireless Environment Sensing}\label{3.1.1}
There are two major challenges in the wireless environment sensing using ISAC-DT enabled IM network.
On the one hand, 
limited wireless resources and severe interference with the large number of IMs bring challenges for low efficiency of spectrum utilization for IM network. 
On the other hand, 
The diverse types of wireless environment sensing information bring challenges for the sensing information transmission and processing.

Facing with these challenges, 
ISAC-DT based wireless environment sensing structurally stored the spectrum sensing information, mines the difference and correlation of spectrum sensing information and extracts the feature information.
Then these feature information can be applied in the 
interference modeling and management, 
channel estimation, 
network optimization, etc.

\subsubsection{Physical Environment Sensing}\label{3.2.1}
In ISAC-DT enabled IM network, there are three main challenges in physical environment sensing.
Firstly, 
the signal optimization \cite{7} in adaptive dynamic physical environments brings challenges for ISAC driven DT.
Secondly,
The diverse sensing information types and large amount of sensing data limits the original sensing information transmission in the IM network.
Thirdly,
In the physical environment imaging of high mobility targets, the accuracy and delay are both strictly restricted, which is a challenge for ISAC-DT enabled IM network.
Finally, 
The large amount of sensing information brings challenges for transmission and processing in the IM network.

Facing with these challenges, 
firstly, we use DT for ISAC signal optimization.
Exploiting the ISAC CSI in 
DT model, the parameters of ISAC signal could be optimized.
Using the prediction of the interference by deep learning, 
effective interference elimination could be performed 
at the receiver.
Secondly, we use DT to reduces the transmitted data 
through semantic feature extraction in 
the sensing information fusion procedure 
of multi-node cooperative sensing.
Thirdly, 
considering the low delay but low accuracy of BS imaging, 
and the high accuracy but high delay of IM imaging, 
the cooperative of BS and IM imaging controlling by DT can meet the strictly restricted physical environment imaging.
Finally, we use DT to 
fuse the diverse sensing information
including the radar point clouds,
radar image, visual image, etc. \cite{9}, 
with the high computing power provided by DT,
realizing accurate sensing of target and environment.

\subsubsection{DT Construction with networked sensing}\label{3.1}
In the networked sensing scenario of IM network, 
multi-BSs cooperatively sense the environment. 
In this case, 
the DT constructed by each BS is independent and isolated.
Therefore,
it is difficult to share the sensing information across DTs, 
which increases the delay and cost in processing and transmission of sensing information.

Facing with these challenges,
we propose a DT construction method with networked sensing using DT split technology.
This technology enables nearby servers with lower computing power 
to construct a local DT model, 
while using high computing power server clusters to construct the global DT model by fusing the local DTs.

\subsection{ISAC-DT based IM Communication}\label{3.3}
In the ISAC-DT enabled IM network, 
the high-dynamic network change in channel information, node distribution, and interference
brings great challenges to the complex process and unfresh data of the IM network 
in physical layer parameters collecting 
and DT updating.
Therefore, 
the optimization effect of channel estimation, multiple access and other processes has been negatively affected.
Facing with the above challenges, we propose the following enabling technologies.

\subsubsection{ISAC-DT based channel 
estimation}\label{3.2.3}
The complex and high-dynamic 
channel of IM network 
bring challenges in channel estimation
with conventional methods. 
Using the sparseness of millimeter-wave (mmWave) channel 
in the angle domain and the close relation between 
the multi-path component and the physical environment, 
the efficient channel estimation algorithm assisted by 
the sensing information in ISAC driven DT 
is developed. 
The learning algorithms in DT algorithm library, 
such as the deep learning based channel estimation algorithm \cite{21} 
is utilized for the localization of 
users and multiple scatterers. 
Then, high-accurate channel estimation method 
is developed based on 
the coupling between the sensing and communication layers.

\subsubsection{ISAC-DT based beam management}

In the era of 6th generation (6G) 
mobile communication system,
the mmWave and Tera Hertz (THz) will be widely applied
in IM network, where the beam 
management faces great challenge
with the high mobility of IMs and 
the highly dynamic environment.
The sensing information 
in ISAC-DT including the location and velocity of IMs 
is applied in beam alignment and tracking \cite{Cui}.
Besides, the sensing information stored in ISAC-DT 
is applied to predict the mobility of IMs, which 
could be applied in predictive beam tracking.

\subsubsection{ISAC-DT based multiple access control}\label{3.3.1}
For the ISAC MAC in large-scale IM network, 
the distribution model of nodes and interference is complex and changing in real-time.
Therefore, the limited resources in IM network cannot support the interaction 
between the network and all terminals 
throughout the entire process 
from initial access to multiple access transmission.

To address this challenge, 
ISAC-DT enabled IM network simplifies, optimizes, and integrates 
the multi access signaling interaction process through DT simulation. 
Furthermore, 
DT can continuously optimize MAC policies 
based on real-time changing nodes and interference distribution, 
ensuring the adaptability of MAC policies to real-time changing environments.

\subsection{ISAC-DT based IM Networking}\label{3.3}
The challenges of ISAC-DT based IM Networking can be summerized as follows.
Firstly, 
the high mobility and multi-domain heterogeneous resources of IMs 
pose challenges to the network management. 
Secondly, 
disguised nodes and malicious attacks emphasize to the security and privacy protection of networking.
Facing with the above challenges, we propose the following enabling technologies.

\subsubsection{ISAC-DT based Neighbor Discovery and Routing}\label{3.3.2}
Due to the high mobility of IM networks 
and the complexity of network topology changes, 
it is necessary to fully consider 
the impact of data transmission and processing delays 
between the DT network topology model and the physical network topology model 
in ISAC-DT enabled IM networks. 
Therefore, it is difficult to synchronize 
the network topology model in DT with the physical network topology model.

Facing with the above challenges, 
ISAC-DT enabled IM network predicts the movement trajectory of IM, 
enables network topology synchronization 
through a combination of DT prediction and low latency ISAC sensing. 
Then DT can provide information for IM neighbor discovery 
through topology prediction.

\subsubsection{DT-based resource management}\label{3.3.3}
In the ISAC-DT enabled IM network, 
network resources, wireless environment sensing resources, and physical environment sensing resources are heterogeneous, 
making it difficult to call heterogeneous data and manage resources \cite{10}. 
This further increases the difficulty of DT construction and updating.

Facing this problem, 
it is necessary to construct multi-dimensional databases and twinning models. 
Among them, 
network data repository and environment data repository are used to store heterogeneous data. 
The basic model constructs network element model, topology model based on the network data repository, 
and environment model models based on the environment data repository.

\subsubsection{DT-based network reliability assurance}\label{3.3.4}
The IM network may encounter disguised nodes and 
malicious attacks during operation. 
Identifying those nodes and attacks is the key 
to ensuring network reliability and data privacy.
ISAC-DT enabled network can identify disguised nodes and malicious attacks 
through the ISAC BSs and DT.
Firstly, 
ISAC BSs monitor and collect the real-time information from IMs,
which can identify potential disguised nodes by monitoring IM information and movement trajectories. 
Secondly, 
BS transmit sensing data to DT,
then DT identifies disguised and attacking nodes 
through data mining and prediction techniques.
Therefore,
through ISAC BSs and DT,
ISAC-DT enabled network can effectively improves the reliability of the IM network.

\section{Use Case}\label{use case}
\begin{figure*}
	\centering
	\includegraphics[width=0.85\linewidth]{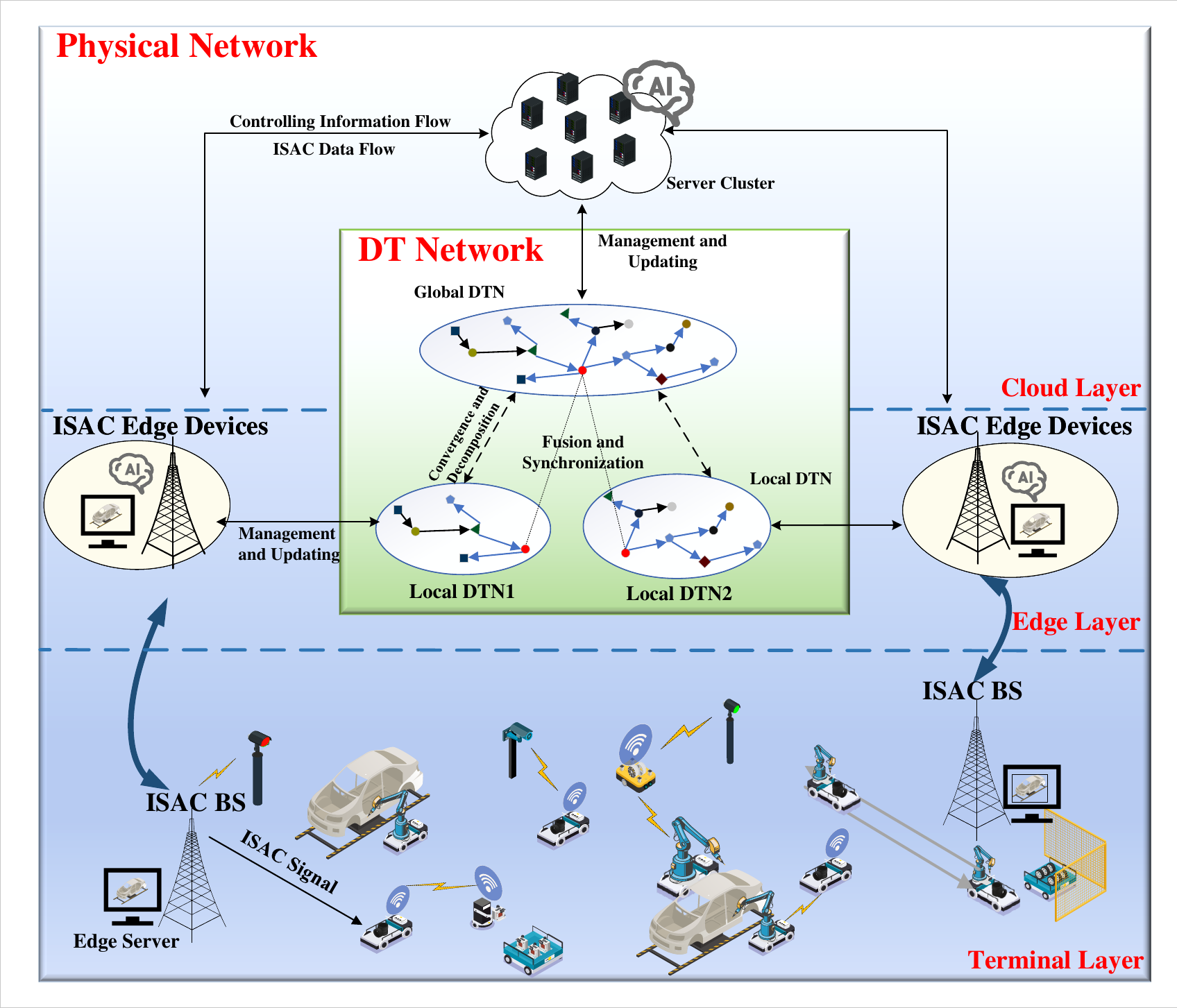}
	\caption{\textcolor{blue}{Use case of ISAC-DT enabled IM network.}}
	\label{fig3}
\end{figure*}

As shown in Fig. \ref{fig3}, 
the ISAC-DT enabled IM network 
is applied in the flexible manufacturing.
In terminal layer, 
the BSs and IMs are empowered with ISAC capability,
which will collect large amount of sensing information.
In edge layer,
MEC servers build the
Local DT using the sensing information from BSs
and IMs. 
In cloud layer, the
global DT is built by aggregating the multiple local DTs
in the edge layer.

The local DT in edge layer will assist the 
optimization of environment sensing,
IM communication, IM networking and the real-time control of IMs
due to the local DT stores the real-time 
sensing information in the physical 
and wireless environment.
The global DT in cloud layer 
will assist the optimization of 
IM networking, IM control and lifecycle management
of flexible manufacturing.

\section{Performance Evaluation}
\label{Simulation Results and Discussion}
In this section, 
the sensing, communication 
networking in ISAC-DT enabled IM network
are simulated. 

\subsection{ISAC-DT based Environment Sensing}
\subsubsection{Cooperative Target Localization}

Fig. \ref{fig_multi_node} shows the performance of
multiple BSs cooperative target localization.
Each BS estimates the location of target.
Then, the estimated locations of multiple BSs are averaged
to obtain the final estimation of target's location.
When the signal-to-noise ratio (SNR) of multiple 
BS are nearly the same, 
multi-BS cooperative sensing has better performance 
compared with single-BS sensing.
However, when the SNRs
of multi-BS are differentiated,
the sensing results of multi-BS need to be 
weighted averaged \cite{multi-node-cooperation}.

\subsubsection{Cooperative Environment Reconstruction}
\begin{figure*}[t]
	\centering
 	\subfloat[]{
		\includegraphics[width=0.45\linewidth]{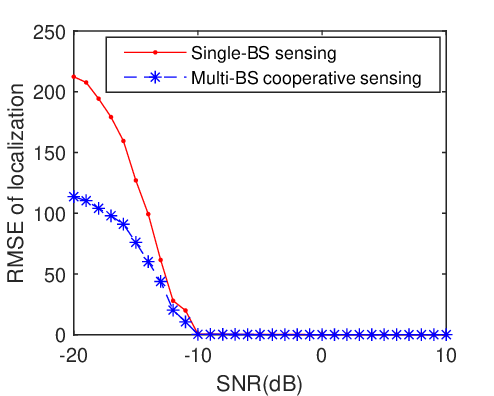} 
		\label{fig_multi_node} 
	}
 	\quad
	\subfloat[]{
		\includegraphics[width=0.47\linewidth]{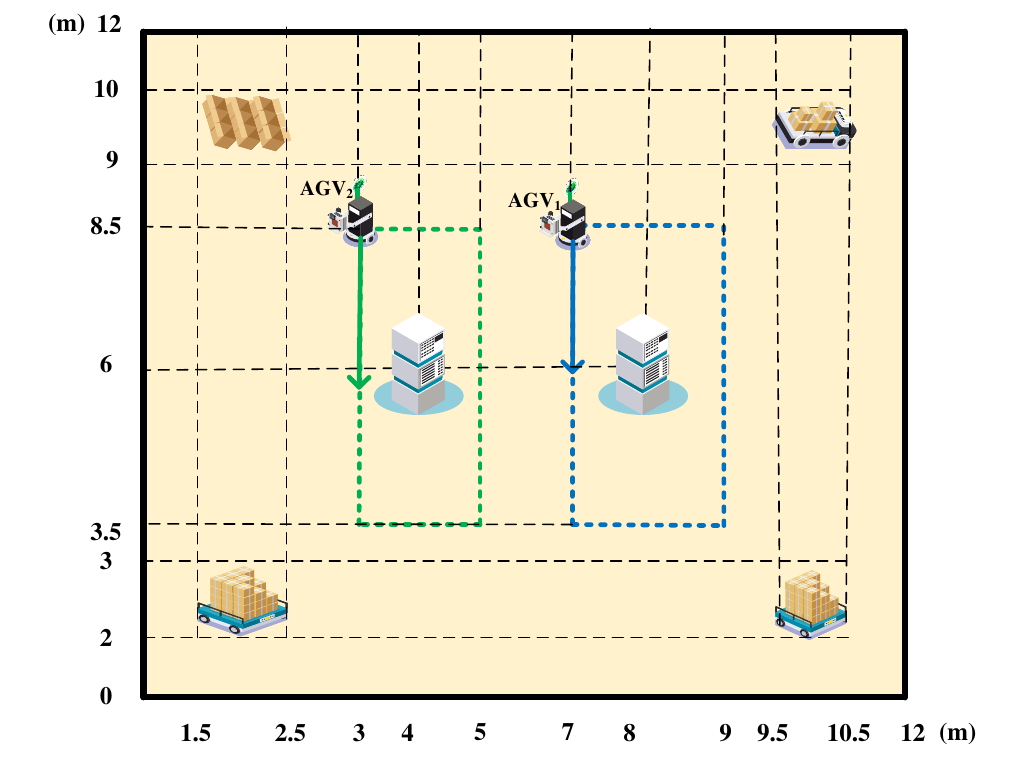}
		\label{fig_imaging_1} 
	}
	\quad  
	\subfloat[]{
		\includegraphics[width=0.47\linewidth]{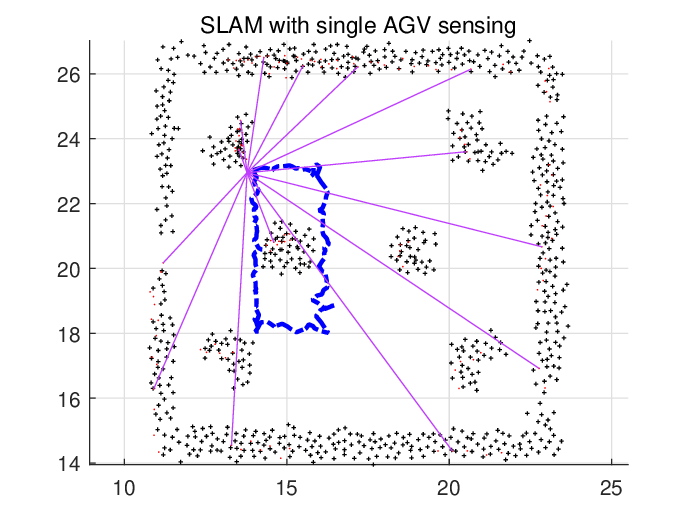} 
		\label{fig_imaging_1} 
	}
        \quad  
	\subfloat[]{
		\includegraphics[width=0.47\linewidth]{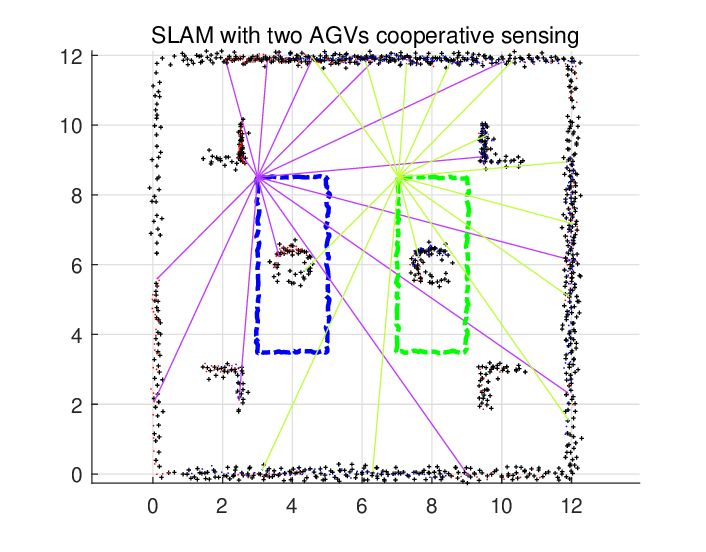} 
		\label{fig_imaging_2} 
	}
	\caption{Cooperative target localization and environment reconstruction.}
	\label{fig4}
\end{figure*}

Fig. \ref{fig_imaging_1} shows the environment of 
a flexible manufacturing workshop.
The Orthogonal Frequency Division Multiplexing (OFDM) 
signal is adopted as ISAC signal in 
environment reconstruction \cite{12}. 
The carrier frequency and bandwidth of ISAC signal 
are set as 28 GHz and 1.23 GHz, respectively.
The AGV with ISAC capability 
performs Simultaneous Localization and Mapping (SLAM).
There are two types of sensing, 
namely active and passive sensing, 
where AGV receives its own or other AGV's
echo signal for environment reconstruction.
Both active and passive sensing are performed 
in an AGV.
Then, the sensing results of active and passive sensing 
are fused to obtain the final sensing results.
Figs. \ref{fig_imaging_1} and \ref{fig_imaging_2}
show the environment reconstruction results 
using single AGV and two AGVs, respectively.
It is obvious that the environment reconstruction
with multiple AGVs cooperative sensing 
has clearer mapping results than that with single AGV sensing,
which is beneficial in building ISAC-DT.

\subsection{ISAC-DT based IM Communication}
Fig. \ref{fig_beam_tracking} reveals the performance of 
ISAC-DT assisted beam tracking,
where the IM is moving towards the BS and 
the beam of BS is tracking the IM.
The traditional beam tracking method is 
the feedback based method, where the feedback from the IM is applied in 
the beam tracking of BS.
Using the assistance of 
the sensing information in ISAC-DT,
the ISAC-DT assisted method is developed,
which improves the spectrum efficiency.
Besides, with the increase of the number of antennas,
the spectrum efficiency is correspondingly improved.

\subsection{ISAC-DT based IM Networking}
\begin{figure*}[t]
	\centering
	\subfloat[]{
		\includegraphics[width=0.56\linewidth]{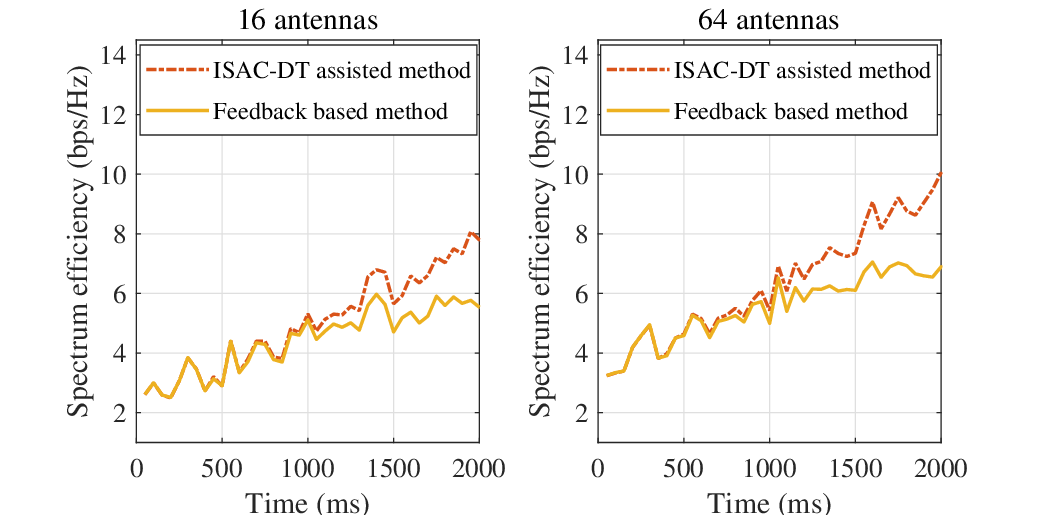}
		\label{fig_beam_tracking} 
	}
	\quad  
	\subfloat[]{
		\includegraphics[width=0.39\linewidth]{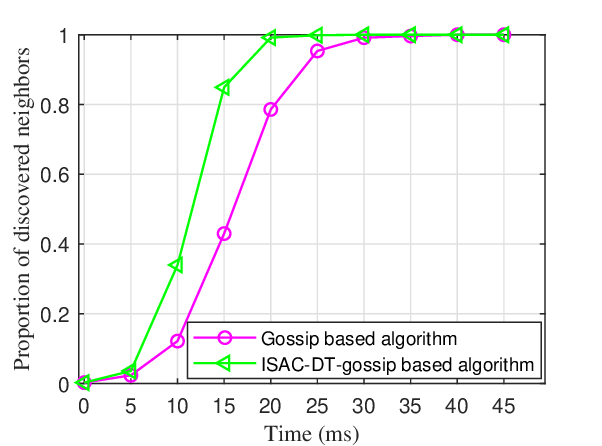} 
		\label{fig6b} 
	}
	\caption{ISAC-DT based beam management and neighbor discovery in IM network.}
	\label{fig6}
\end{figure*}

Fig. \ref{fig6b} illustrates the relation between 
the proportion of discovered neighbors and 
the time of neighbor discovery
with the most efficient algorithm,
namely the gossip based algorithm \cite{15}. 
Using the ISAC-DT-gossip based algorithm,
where the sensing information including the distribution of IMs
is applied to improve the efficiency of gossip based algorithm, 
the speed of neighbor discovery is improved.

\section{Conclusion}\label{Conclusion}
IMs including industrial machines, 
unmanned aerial vehicles (UAVs), and unmanned vehicles
are continuously reconfiguring the traditional industry
with the assistance of IM network.
This article proposes the
ISAC driven DT for IM network,
and its architecture and enabling technologies are provided.
With ISAC technology, 
IM network is empowered with ubiquitous sensing capabilities
and the sensing information is structurally
stored in the ISAC-DT, 
which is further applied In the optimization of sensing, 
communication, networking and control schemes of IMs.
The use case of ISAC-DT enabled IM network is revealed
in the scenario of flexible manufacturing.
The ISAC-DT has great potential in improving the efficiency of
sensing, communication and control of IM network,
ushering the era of the symbiosis of human and IMs.

\bibliographystyle{IEEEtran}
\bibliography{reference}

\end{document}